\documentclass{article}
\usepackage{csquotes}
\usepackage{spconf,amsmath,graphicx,siunitx,microtype,amsfonts}
\usepackage[numbers]{natbib}
\usepackage{array}
\usepackage{booktabs}
\usepackage{microtype}
\usepackage{url}
\usepackage{hyperref}
\usepackage[all]{hypcap}
\usepackage{amsmath}
\usepackage{siunitx}
\usepackage{color}
\usepackage{multirow}
\usepackage{booktabs}
\usepackage{amsmath,graphicx,url,times,booktabs,tabularx,amsfonts,siunitx}
\usepackage{algorithmic}
\usepackage{graphicx}
\usepackage{textcomp}
\usepackage{xcolor}
\usepackage{latexsym}
\usepackage{bbding}
\usepackage{footnote}
\ninept

\def\L{{\cal L}}
\title{Retrieval-Augmented Text-to-Audio Generation}

\name{
      Yi Yuan,
      Haohe Liu,
      Xubo Liu,
      Qiushi Huang,
      Mark D. Plumbley,
      Wenwu Wang
      }
      
\address{School of Computer Science and Electronic Engineering, University of Surrey, UK\\
 }

\begin{document}

\maketitle

\begin{abstract}
Despite recent progress in text-to-audio (TTA) generation, we show that the state-of-the-art models, such as AudioLDM, trained on datasets with an imbalanced class distribution, such as AudioCaps, are biased in their generation performance. Specifically, they excel in generating common audio classes while underperforming in the rare ones, thus degrading the overall generation performance. We refer to this problem as long-tailed text-to-audio generation. To address this issue, we propose a simple retrieval-augmented approach for TTA models. Specifically, given an input text prompt, we first leverage a Contrastive Language Audio Pretraining (CLAP) model to retrieve relevant text-audio pairs. The features of the retrieved audio-text data are then used as additional conditions to guide the learning of TTA models. We enhance AudioLDM with our proposed approach and denote the resulting augmented system as Re-AudioLDM. On the AudioCaps dataset, Re-AudioLDM achieves a state-of-the-art Frechet Audio Distance (FAD) of $1.37$, outperforming the existing approaches by a large margin. Furthermore, we show that Re-AudioLDM can generate realistic audio for complex scenes, rare audio classes, and even unseen audio types, indicating its potential in TTA tasks.
\end{abstract}

\begin{keywords}
	Audio generation, retrieval-information, diffusion model, deep learning, long tail problem
\end{keywords}

\section{Introduction}
\label{sec:intro}
The landscape of text-to-audio (TTA) generation has been revolutionized by advancements in diffusion-based generative modelling~\cite{audioldm,makeaudio,wavjourney}. Leveraging powerful backbone models such as CLAP~\cite{audioldm} and large language model~(LLM)~\cite{tango}, these models are capable of extracting semantic information and enabling the creation of high-fidelity audio from textual descriptions. 

\begin{figure}[t]
    \centering
    \includegraphics[width=\linewidth]{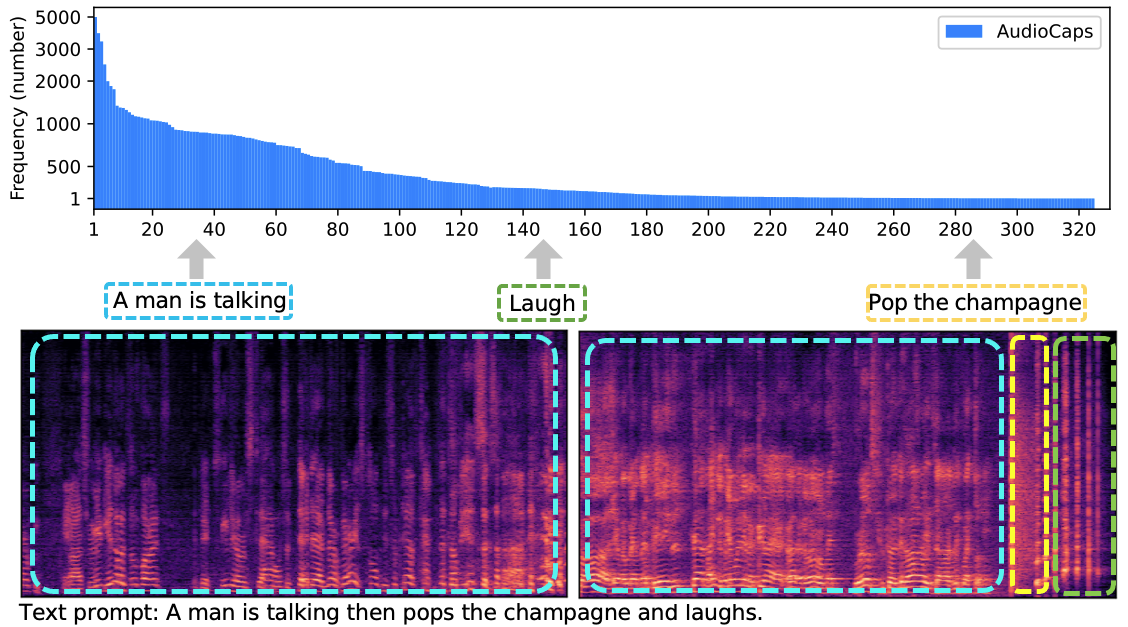}
    \vspace{-6mm}
    \caption{The long-tailed problem in AudioCaps dataset (top). Example audio clips (bottom) generated with the baseline model~(left) and Re-AudioLDM~(right). }
    \label{frequency}
    \vspace{-1em}
\end{figure}

In this work, we show that due to the scarcity and diversity of audio training data, bias appears in these state-of-the-art models, leading to significant performance degradation. Figure~\ref{frequency} (top) draws a statistical analysis conducted on the $327$ labels of AudioCaps~\cite{audiocaps}, one of the largest audio-text datasets, indicating a notable imbalance in data distribution. The bottom-left graph of Figure~\ref{frequency} shows a sample result of the state-of-the-art model trained with AudioCaps, when giving the prompt ``\textit{A man is talking then pops the champagne and laughs}'', the model could only generate the content for ``\textit{man talking}'', but miss uncommon or complex events such as``\textit{champagne popped}'' then followed by ``\textit{laugh}''. Hence, an inherent limitation is seen due to the constrained scope and variability of the training dataset, where the quality of generated sounds seems largely correlated with their frequency of appearance during training. In this regard, these models can faithfully generate realistic audio clips for common sound events, but they may generate incorrect or irrelevant audio clips when encountering less frequent or unseen sound events. 

We denote this the \textit{long-tailed text-to-audio generation} problem, which influences the model performance in diversity and restricts the applicability of these models, especially in real-world scenarios. 
Our motivation is to develop a robust TTA framework that breaks the barrier of imbalanced data and achieves realistic generation on diverse acoustic entities.

We propose a novel retrieval-augmented TTA framework to address the long-tailed generation issue. We enhance the state-of-the-art TTA model, AudioLDM~\cite{audioldm}, with a retrieval module, dubbed Re-AudioLDM.
Specifically, we first use the input text prompt to retrieve relevant references (e.g., text-audio pairs) from dataset~(e.g., AudioCaps), and then use a pre-trained audio model and a language model to extract the acoustic and textual features, respectively. These extracted features are then further given to the cross-attention~\cite{transformer} module of the LDM to guide the generation process. The retrieved audio-text pairs serve as supplementary information that helps improve the modelling of low-frequency audio events in the training stage. In the inference stage, the retrieval-augmented strategy also provides references in relation to the text prompt, ensuring a more accurate and faithful audio generation result. 

We perform extensive experiments on events with different frequencies of occurrence in the dataset. We show that Re-AudioLDM provides a stable performance among a variety of audio entities. It significantly improves the performance for tail classes over the baseline models, demonstrating that it can provide effective alleviation for long-tail TTA issues. Furthermore, as compared with the baseline models, Re-AudioLDM is capable of generating more realistic and complex audio clips, including rare, complex, or even unseen audio events. As the example with the same prompt shown in Figure~\ref{frequency} (bottom), where Re-AudioLDM~(bottom-right) can generate both uncommon entities ``\textit{champagne popped}'' with a complex structure followed with the sound of ``\textit{laugh}'', achieving a better result than the baseline models with all the required entities and semantic orders correctly. In addition, Re-AudioLDM achieves an FAD score of $1.37$, outperforming state-of-the-art TTA models by a large margin.

\begin{figure*}[htbp]
    \centering
    \includegraphics[width=0.9\linewidth]{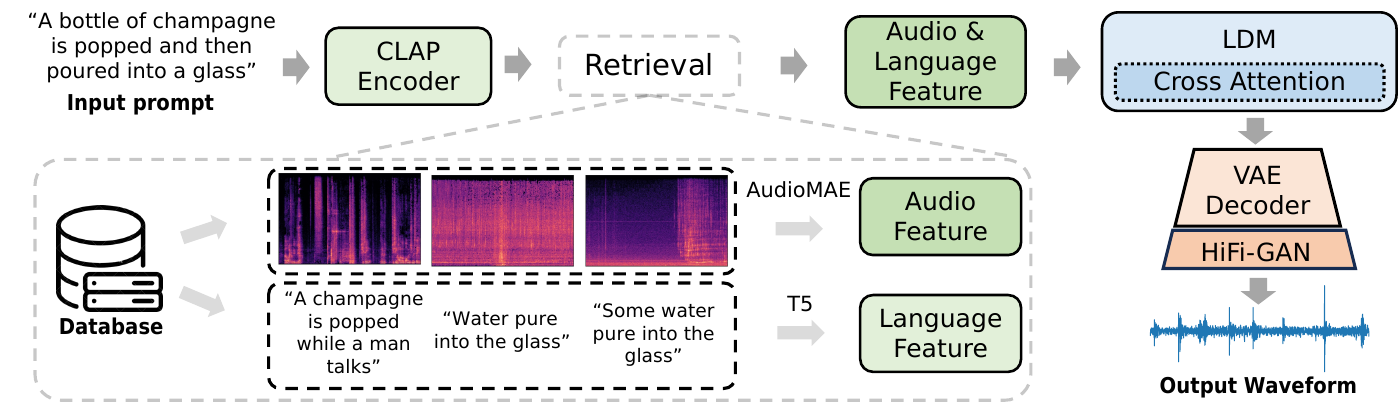}
    \caption{The overview structure of Re-AudioLDM}
    \label{fig:overview}
\end{figure*}

The remainder of this paper is organized as follows. Section~\ref{sec:related_works} introduces the related works of audio generation and retrieval-based models, followed by the details of Re-AudioLDM in Section~\ref{sec:method}. Section \ref{sec:exp} presents the experimental setting and Section \ref{sec:result} shows the results and ablation studies. Conclusions are given in Section \ref{sec:conclusion}.

\section{Related work}
\label{sec:related_works}
Our work relates to two major works, diffusion-based text-to-audio models and retrieval-based generative models. These two fields are briefly discussed in the following subsections.

\subsection{Audio Generation}  
Recent works on audio generation follow an encoder-decoder framework~\cite{audioldm,leveraging}. The model first uses an encoder to encode the information into a latent representation, which can be decompressed into a mel-spectrogram feature. The decoder then uses a variational autoencoder~(VAE) and a generative adversarial network~(GAN) vocoder to turn such features into waveforms. Liu \textit{et al.}~\cite{Liu-tts} has used PixelSNAIL~\cite{pixelsnail} as the encoder to represent labels while Iashin and Rahtu ~\cite{specvqgan} applied a GPT2~\cite{gpt2} as the encoder to encode input images. Subsequently, diffusion-based models have been used for latent token generation. Yang \textit{et al.}~\cite{diffsound} replaces the transformer-based encoder with a diffusion-based encoder. Liu \textit{et al.}~\cite{audioldm} uses the CLAP model~\cite{clip} to obtain embeddings for the input data (audio or text), and uses the Latent Diffusion Model~(LDM) as the token generator. Ghosal \textit{et al.}~\cite{tango} then further improves this framework by replacing CLAP with LLM~\cite{t5}. 

\subsection{Retrieved Information Aggregation}
Several studies in the field of image generation have considered leveraging retrieved information. Li \textit{et al.}~\cite{Li2022} extract image features from a training set, and place them in a memory bank which is then used as a parallel input condition for audio generation. Blattmannet \textit{et al.}~\cite{blattmann2022} present a nearest-neighbours strategy to select related image samples from a neighbourhood area. The KNN-Diffusion~\cite{sheynin2022} uses image features obtained from large-scale retrieval databases during the inference stage to perform new-domain image generation. Chen \textit{et al.}~\cite{reimagen} extend the image-only retrieval into image-text pair retrieval, augmenting both high-level semantics and low-level visual information for a diffusion model. In contrast, no similar works have been done for audio generation, and Re-AudioLDM is the first attempt to introduce retrieved information from a dataset to improve the text-to-audio generation performance. 

\section{Proposed Method}
\label{sec:method}

Similar to previous audio generation works~\cite{audioldm,tango,tuning}, Re-AudioLDM is a cascaded model including three parts: input embedding, diffusion-based feature generator, and a pipeline to reconstruct the waveform from the latent feature. 
\subsection{Text and Retrieval Embedding Encoder}
\label{sec:encoder}
Re-AudioLDM takes two paralleled inputs: a text input $c_t$ as low-level semantic information, and a set of text-audio pairs as retrieval augmentation $c_r$ for high-level semantic-audio information. The text embedding $\boldsymbol{E}^{t}$ is obtained as: 
\begin{equation}
\label{eqa:1}
    \boldsymbol{E}^{t} = \textit{f}_{\text{clap}}(c_t)
\end{equation}
which $\textit{f}_{\text{clap}}(\cdot)$ is the CLAP model \cite{clap} used for text encoding, as in AudioLDM~\cite{audioldm}. The retrieved information $c_r =[<\text{text}_1,\text{audio}_1>,<\text{text}_2,\text{audio}_2>,...,<\text{text}_k,\text{audio}_k>] $ are the top-k neighbours selected through the similarity comparison between the embedding of the target caption and those of the retrieval dataset. Here for each pair, the multi-modal embedding is divided into two groups of concatenation, presented as audio retrieval $\boldsymbol{E}^{ra}$ and text retrieval $\boldsymbol{E}^{rt}$, encoded as:
\begin{align}
\label{eqa:2}
\boldsymbol{E}^{ra} = \text{CAT}[\textit{f}_{\text{mae}}(\text{audio}_1), ... , \textit{f}_{\text{mae}}(\text{audio}_k)],\\
\label{eqa:3}
\boldsymbol{E}^{rt} = \text{CAT}[\textit{f}_{\text{t5}}(\text{text}_1), ... ,\textit{f}_{\text{t5}}(\text{text}_k)]
\end{align}

\noindent
where $\textit{f}_{\text{t5}}(\cdot)$ is a pre-trained T5 model~\cite{t5} for obtaining the text embedding, and $\textit{f}_{\text{mae}}(\cdot)$ is a pre-trained AudioMAE model~\cite{mae} for obtaining the embedding of the paired audio. 
\subsection{Retrieval-Augmented Diffusion Generator}
Re-AudioLDM uses LDM as the generator to obtain the intermediate latent token of the target audio. The diffusion model involves two processes, a forward process to gradually add noise into the latent vectors and a reverse process to progressively predict the transition noise of the latent vector in each step. During the forward step, the latent representation $z_0$ is transformed into a standard Gaussian distribution $z_n$ with a continuous noise injection:
\begin{align}
q(\boldsymbol{z}_{n}|\boldsymbol{z}_{n-1})&=\mathcal{N}(\boldsymbol{z}_{n};\sqrt{1-\beta_{n}}\boldsymbol{z}_{n-1},\beta_{n}\boldsymbol{I}), \\
\label{forwardprocess}
q(\boldsymbol{z}_{n}|\boldsymbol{z}_{0})&=\mathcal  N(\boldsymbol{z}_{n};\sqrt{\bar{\alpha}_{n}}\boldsymbol{z}_{0},(1-\bar{\alpha}_{n})\boldsymbol{\epsilon})
\end{align} 
\noindent
where $\epsilon$ denotes the Gaussian noise with ${\alpha}_{n} = 1 -\beta_{n}$ controlling the noise level. 
In the reverse process, LDM learns to estimate the distribution of noise $\boldsymbol{\epsilon}_{\theta}$ in the latent space, given conditions from the text embedding $\boldsymbol{E}^{t}$, calculated with equation~(\ref{eqa:1}), and the retrieved embedding $\boldsymbol{E}^{ra}$ and $\boldsymbol{E}^{rt}$, calculated with equation~(\ref{eqa:2}) and ~(\ref{eqa:3}), respectively.
The LDM model applies UNet as the general structure, where the input layer takes the noisy latent vector $\boldsymbol{z}_{n}$, text embedding $\boldsymbol{E}^{t}$, and the time step $\textit{n}$ as the condition. Then the retrieved information of both text and audio is shared with all the cross-attention blocks within the remaining layers. Employing a re-weighted training objective ~\cite{DDPM}, LDM is trained by:
\begin{equation}
\label{trainingobjective}
L_{n}(\theta)=\mathbb{E}_{\boldsymbol{z}_{0},\boldsymbol{\epsilon},n}\left \| \boldsymbol{\epsilon} - \boldsymbol{\epsilon}_{\theta}(\boldsymbol{z}_{n},n,\boldsymbol{E}^{t},\text{Attn}(\boldsymbol{E}^{ra},\boldsymbol{E}^{rt})) \right\|^2_{2}
\end{equation}

\subsection{VAE Decoder \& Hifi-GAN Vocoder}
\label{sec:pipline}
Re-AudioLDM utilizes a combination of a VAE and a HiFi-GAN as the general pipeline for reconstructing waveform from the latent feature tokens. During the training stage, VAE learns to encode the mel-spectrogram into the intermediate representation and then decode it back to mel-spectrogram, while Hifi-GAN is trained to convert mel-spectrogram into waveform. For inference, Re-AudioLDM applies the VAE decoder for mel-spectrogram reconstruction and HiFi-GAN for waveform generation. 


\begin{table*}[htbp]
\label{tab:result}
\centering
\small
\begin{tabular}{ccccccccc}
\toprule
 Model  & Dataset & Retrieval Info & Retrieval Number  & KL $\downarrow$ &  IS $\uparrow$& FAD $\downarrow$ & $\operatorname{CLAP}_{\operatorname{score}}$(\%)$\uparrow$ \\
\midrule
AudioGen~\cite{audiogen}
                       &AC+AS+8 others&\XSolidBrush&\XSolidBrush &    $1.69$    &  $5.13$   &  $2.15$   & $23.44$ \\
\midrule
AudioLDM~\cite{audioldm}
                       &AC+AS+2 others&\XSolidBrush&\XSolidBrush &    $1.66$    &  $6.51$   &  $2.08$   & $25.39$ \\
\midrule
Tango~\cite{tango}
                       &AudioCaps&\XSolidBrush&\XSolidBrush &   $1.32$     &  $6.45$   &  $1.68$  & $29.28$ \\
\midrule
\multirow{4}{*}{Re-AudioLDM-S}
                       &AudioCaps &\XSolidBrush&  \XSolidBrush   &  $1.63$   &  $6.48$   &   $2.31$    & $26.75$ \\
                       &AudioCaps &Audio&   $3$     &  $1.54$    &  $6.88$   &  $1.95$     & $31.05$ \\
                       &AudioCaps & Audio \& Text &  $3$     &  $1.27$   &  $7.31$   &   $1.48$    & $37.07$ \\
                       &AudioCaps & Audio \& Text&   $10$     &  $1.23$   &  $7.33$   &   $1.40$    & $\mathbf{37.15}$ \\
\midrule
Re-AudioLDM-L
                       &AudioCaps&Audio \& Text& $10$ &   $\mathbf{1.20}$     &  $\mathbf{7.39}$   &  $\mathbf{1.37}$  & $37.12$ \\
\bottomrule
\end{tabular}
\label{tab:pre-train}
\caption{The comparison between different frameworks, with and without retrieval information. AC and AS are short for AudioCaps~\cite{audiocaps} and AudioSet~\cite{audioset} respectively.}
\end{table*}

\section{Experiments}
\label{sec:exp}

\subsection{Datasets}
We use the AudioCaps dataset~\cite{tau2019} for the experiments, which comprises $46,000$ ten-second audio clips, each paired with a human-annotated caption. We follow the official training-testing split, where each training audio clip is assigned a single caption, while in the testing split, each audio clip is annotated with five captions. During the inference stage, we employ the first caption of each audio clip that appears in the test split as the text input. The remaining four captions are used only for the ablation study in Section~\ref{sec:result}.

\subsection{Experiment Setup}

\textbf{Data Preparation.}
For a retrieval-based AudioCaps dataset, we apply a CLAP-score based retrieval function to find the top-50 nearest neighbours of the target text embedding. The waveform and the text from each neighbour are stored as a text-audio pair. It is noted that for both training and testing samples, the target sample is excluded from the retrieval information, which can avoid any access to the target data during both the training and inferencing stages.

\noindent
\textbf{Implementation Detail.}
As a cascaded model, the encoder and decoder parts of Re-AudioLDM are trained separately with audio clips sampled at $16$ kHz. For the target, we use the short-time Fourier transform~(STFT) with a window of 1024 samples and a hop size of 160 samples, resulting in a mel-spectrogram with 64 mel-filterbanks. Then, a VAE model is applied to compress the spectrogram with a ratio of $4$, resulting in a feature vector with a frequency dimension of 16. For the information provided by the retrieval strategy, the text feature is directly extracted by a pre-trained T5-medium model, presenting a fixed sequence length of 50. The audio feature, on the other hand, is first converted into filter banks with 128 mel-bins and then processed by a pre-trained AudioMAE model, leading to a vector of dimension 32. 

\noindent
\textbf{Training Detail.}
The LDM is optimized with a learning rate of $5.0\times10^{-5}$. Re-AudioLDM is trained for up to $80$ epochs with a batch size of $4$ and the evaluation is carried out every $100,000$ steps. Re-AudioLDM-S applies a UNet architecture consisting of $128$ channels, while we enlarge the model into Re-AudioLDM-L with $196$ channels for experiments on more complex models. 

\noindent
\textbf{Evaluation Metrics.}
Following Liu \textit{et al.}, we use the Inception Score~(IS), Fréchet Audio Distance~(FAD), and Kullback–Leibler~(KL) divergence to evaluate the performance of Re-AudioLDM. A higher IS score indicates a larger variety in the generated audio, while lower KL and FAD scores indicate better audio quality. For the semantic-level evaluation, we calculate the cosine similarity between the output audio embedding and the target text embedding calculated by the CLAP encoders, which demonstrates the correlation between audio and text.

\section{Results}
\label{sec:result}

\subsection{Evaluation Results}
The experiments are carried out on AudioCaps evaluation set. We compare the performance with several state-of-the-art frameworks, including AudioGen~\cite{audiogen}, AudioLDM~\cite{audioldm} and Tango~\cite{tango}. Selecting only the first caption of each audio clip as the text description, each framework infers $975$ 10-second audio clips with a sampling rate of $16$ kHz. Table~\ref{tab:result} compares the metrics achieved with different text-to-audio models, where Re-AudioLDM outperforms other methods by a large margin. It is noted that without the additional information provided by retrieval, Re-AudioLDM does not exhibit any advantage and is generally inferior to Tango, the current state-of-the-art model on AudioCaps. However, upon incorporating retrieval information, Re-AudioLDM successfully outperformed the baseline models in all four evaluation metrics. By enlarging the size of the hidden layer in LDM structure, the Re-AudioLDM-L using 10 retrieved pairs further decreases the FAD score to below $1.4$, which is a significant improvement over the baseline frameworks. 

\subsection{Ablation Study}
\label{ablation}
\textbf{Retrieval Type.}
Experiments in Table~\ref{tab:result} show the results on different retrieval information, e.g. audio, text, or neither. With the audio features extracted by AudioMAE, only a slight improvement is achieved by Re-AudioLDM, mainly because Re-AudioLDM misses the relationship between sound events, although it captures the features of related sound events. By adding the paired text information of each retrieved audio clip, Re-AudioLDM learns the relationship between audio features and high-level semantic information, contributing a significant improvement in capturing highly related semantic features for the sound events.

\begin{figure}[htbp]
    \centering
    \includegraphics[width=\linewidth]{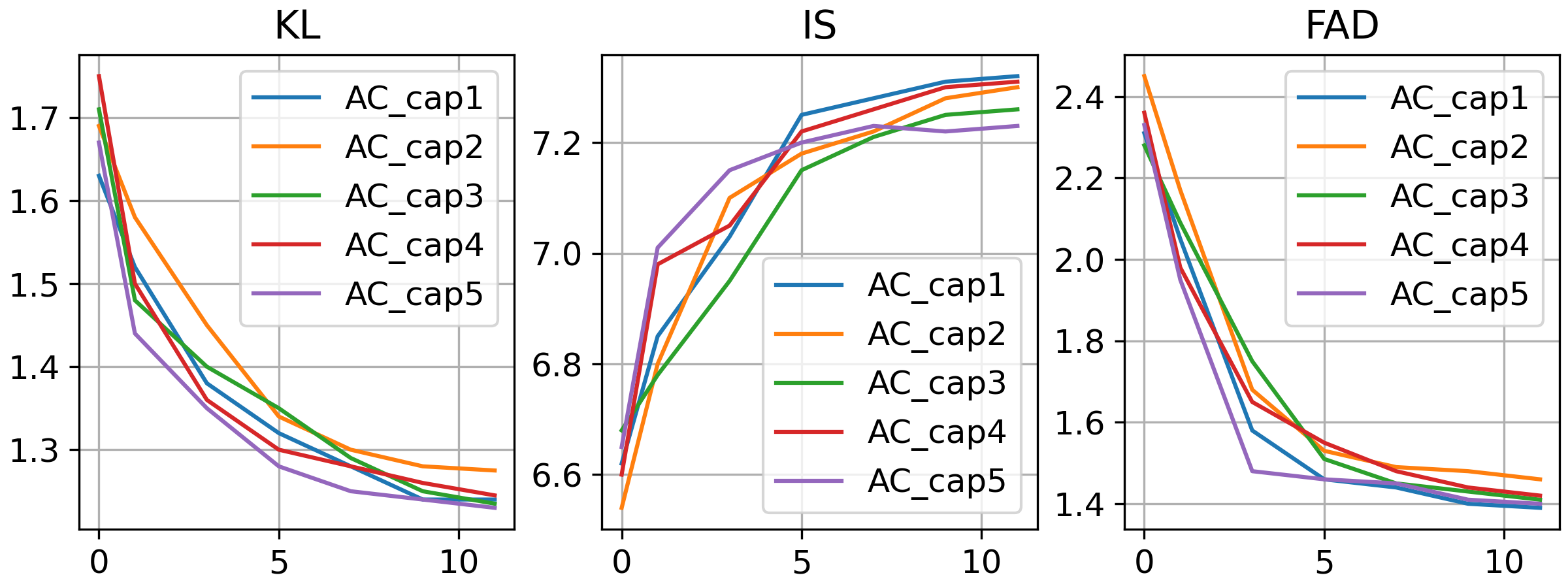}
    \caption{Performance comparison on numbers of retrieved information, where AC\_cap 1-5 refers to the caption groups of the testing set. }
    \label{fig:retriva-number}
\end{figure}

\noindent
\textbf{Number of Retrieved Pairs.}
Several experiments are carried out to assess the impact of the number of retrieved audio-text pairs on audio generation performance. As depicted in Figure~\ref{fig:retriva-number}, the incorporation of retrieval information improves the performance, as the number of retrieved pairs increases, while such improvement slows down after the number reaches five and it becomes flattened at around ten. Therefore, in order to strike a balance between training costs and model performance, the number of retrieved pairs is chosen empirically to be in the range of $3$ to $5$ for this data.

\begin{figure}[htbp]
    \centering
    \includegraphics[width=\linewidth]{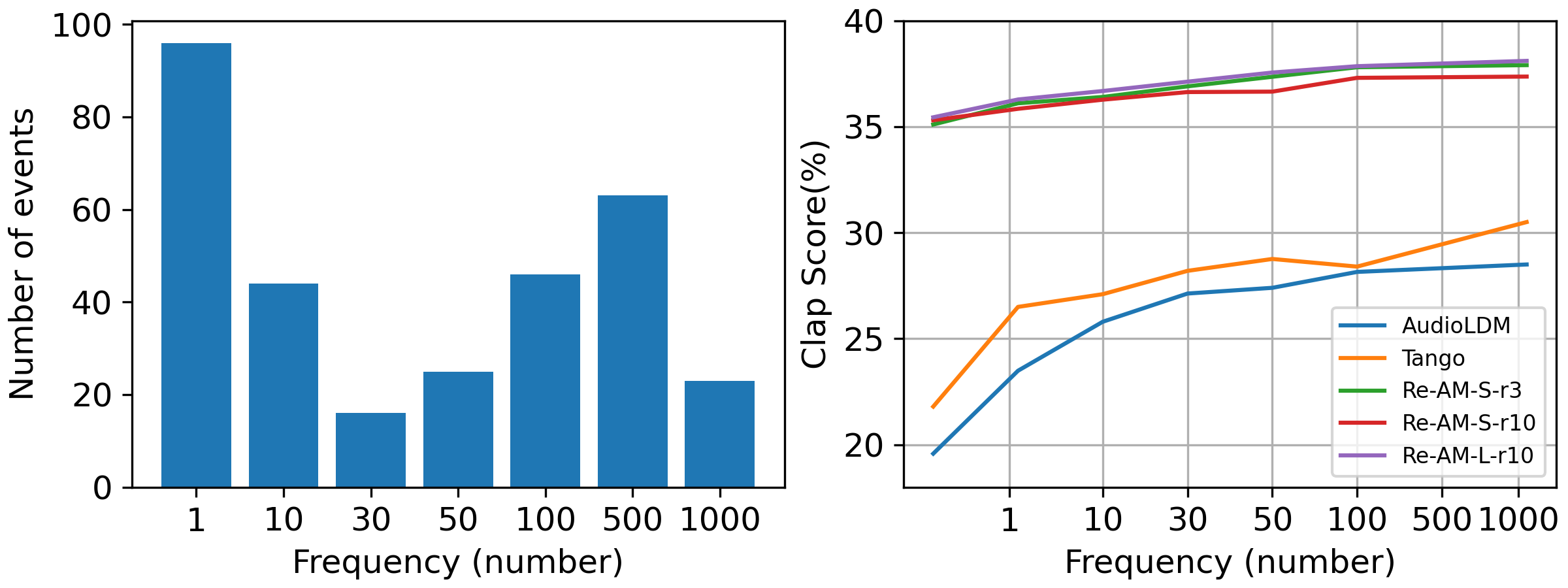}
    \caption{Performance on different frequency entities, where S and L indicate model size and r refers to the number of retrieved clips. }
    \label{fig:retrival-freq}

\end{figure}

\noindent
\textbf{Long-Tailed Situations.}
Re-AudioLDM aims to tackle the long-tailed generation problem and generate more realistic audio clips on uncommon or unseen sound events. In order to evaluate the accuracy of each generated audio clip, we applied the CLAP score~\cite{clap} to show the relationship between the audio clip and text description. We first calculate the frequency of the occurrence of each sound event by counting the label of each audio clip and then illustrate the model performance by averaging the CLAP score of each sound class for the AudioCaps testing set. The bar chart on the left side of Figure~\ref{fig:retrival-freq} presents a statistical analysis of the quantities of all $327$ sound event classes in the AudioCaps training set. Similar to Figure~\ref{frequency} (top), tail classes constitute a significant portion, especially in the label group of 1 and 10. Figure~\ref{fig:retrival-freq} (right) shows the performance of each model on the events with different frequencies of event occurrence within the training set. Despite the initial gap in highly frequent audio events between Re-AudioLDM and baseline models, the baseline models perform worse when handling tailed entities. However, Re-AudioLDM has demonstrated more stable results, with a decrease of less than $3$ in the CLAP score as the frequency of event occurrence is reduced in training data. Hence, Re-AudioLDM can reduce the degradation of output quality when generating tailed sound events, enhancing the overall model performance. 

\noindent
\textbf{Zero-Shot Generation.}
For experiments on unseen entities, we evaluate several scenarios with events that are excluded during training. In Figure~\ref{fig:retrival-freq} (right), we found that baseline models show performance degradation on generating unseen audio~(zero frequency occurrence). This may be because the model has not learned the features of unseen entities, while Re-AudioLDM can still achieve realistic results by providing related audio and semantic information. Hence, with essential retrieval information, Re-AudioLDM has the potential to generate sounds which are excluded from training data. The retrieval-based generation may significantly enhance the robustness of zero-shot tasks, which is one of the directions we will explore in the future. 

\noindent
\textbf{Comparison with Mixup Strategy.}
Another way to address the class imbalance problem is to use a mixup strategy \cite{kong2020panns}. While mixup can increase the occurrence frequency for the tail entities, it also introduces more complex audio examples, as well as the synthetic audio data that may not align with real-world distributions. The results in~\cite{audioldm} have shown that the mixup strategy leads to degradation in overall performance. In contrast to the mixup method, our proposed retrieval augmentation strategy reduces the complexity of the training processes, resulting in an overall performance improvement.

\section{Conclusion}
\label{sec:conclusion}
In this paper, we have presented a retrieval-augmented model, Re-AudioLDM, to tackle the long-tailed problem in AudioCaps. The comparisons with current state-of-the-art models (i.e., AudioLDM and Tango) using several performance metrics (i.e., FAD, and CLAP-score) demonstrate that Re-AudioLDM can significantly enhance the performance of TTA models in generating high-fidelity audio clips. By integrating retrieved features, Re-AudioLDM not only achieves improvements in overall performance, but enables the generation of rare or unseen sound entities. In future work, we will investigate the model with external large datasets and explore the potential of the model in downstream tasks, such as zero-shot generation.

\section{ACKNOWLEDGMENT}
\label{sec:ack}
This research was partly supported by a research scholarship from the China Scholarship Council~(CSC), funded by British Broadcasting Corporation Research and Development~(BBC R\&D), Engineering and Physical Sciences Research Council~(EPSRC) Grant EP/T019751/1 ``AI for Sound'', and a PhD scholarship from the Centre for Vision, Speech and Signal Processing~(CVSSP), University of Surrey. 
For the purpose of open access, the authors have applied a Creative Commons Attribution~(CC BY) license to any Author Accepted Manuscript version arising.
\bibliographystyle{IEEEtran}



\end{document}